\pdfoutput=1

\documentclass[10pt,letterpaper]{article}
\usepackage[T1]{fontenc}
\usepackage{inputenc}
\usepackage[english]{babel}
\usepackage[nohead,includefoot,margin=2cm]{geometry}
\usepackage[compact,raggedright,small]{titlesec}
\usepackage[affil-it]{authblk}
\usepackage[runin]{abstract}
\usepackage{multicol}
\usepackage{blindtext}
\usepackage{upgreek}
\usepackage{graphicx}
\usepackage{float}
\usepackage{hyperref}
\usepackage{titlesec}
\usepackage{amsmath}
\usepackage{nameref}
\usepackage[superscript,biblabel,compress]{cite}
\usepackage[shortcuts]{extdash}
	 
\addto{\Affilfont}{\small}

\title{\large\bfseries Multiheterodyne spectroscopy using interband cascade lasers}
\date{\small (Dated: September 6, 2017)}
\author[1,3,$\dagger$]{L.~A.~Sterczewski}
\author[1,$\dagger$]{J.~Westberg}
\author[1]{L.~Patrick}
\author[2]{C.~S.~Kim}
\author[4]{M.~Kim}
\author[2]{C.~L.~Canedy}
\author[2]{W.~W.~Bewley}
\author[2]{C.~D.~Meritt}
\author[2]{I.~Vurgaftman}
\author[2]{J.~R.~Meyer}
\author[1,*]{G.~Wysocki}

\affil[1]{Department of Electrical Engineering, Princeton University, Princeton, New Jersey 08544, USA}
\affil[2]{Naval Research Laboratory, Code 5613, Washington, DC 20375, USA}
\affil[3]{Faculty of Electronics, Wroclaw University of Science and Technology, Wroclaw 50370, Poland}
\affil[4]{Sotera Defense Solutions, Inc., 7230 Lee DeForest Drive, Suite 100, Columbia MD 21046, USA}
\affil[ ]{}
\affil[$\dagger$]{These authors contributed equally to this work.}
\affil[*]{e-mail: gwysocki@princeton.edu}

\begin{document}
		
   \maketitle
 
   \begin{abstract}

     \noindent While mid-infrared radiation can be used to identify and quantify numerous chemical species, contemporary broadband mid-IR spectroscopic systems are often hindered by large footprints, moving parts and high power consumption. In this work, we demonstrate multiheterodyne spectroscopy using interband cascade lasers, which combines broadband spectral coverage with high spectral resolution and energy-efficient operation. The lasers generate up to 30~mW of continuous wave optical power while consuming less than 0.5~W of electrical power. A~computational phase and timing correction algorithm is used to obtain kHz linewidths of the multiheterodyne beat notes and up to 30~dB improvement in signal-to-noise ratio. The versatility of the multiheterodyne technique is demonstrated by performing both rapidly swept absorption and dispersion spectroscopic assessments of~low-pressure ethylene (C$_2$H$_4$) acquired by extracting a single beat note from the multiheterodyne signal, as well as broadband multiheterodyne spectroscopy of methane (CH$_4$) acquired with all available beat notes with microsecond temporal resolution and an instantaneous optical bandwidth of 240~GHz. The technology shows excellent potential for portable and high-resolution solid state spectroscopic chemical sensors operating in the mid-infrared.\\
     %
     
   \end{abstract}

   \begin{multicols}{2}
     
     \section{Introduction}

     The interband cascade laser~\cite{yang_infrared_1995,meyer_type-ii_1996} (ICL) is often referred to as a hybrid between a conventional diode laser, that generates light from electron-hole recombination, and an intersubband quantum cascade laser~\cite{faist_quantum_1994} (QCL), that stacks multiple stages for efficient photon generation. Its low threshold drive power (29~mW demonstrated for cw operation at room-temperature)~\cite{vurgaftman_rebalancing_2011} and high wall-plug efficiency ($\leq$18\%)~\cite{vurgaftman2015interband} make it particularly well-suited for power-constrained sensing applications; e.g., battery operated devices. NASA's selection of a single-mode ICL for methane sensing on its Mars Curiosity Mission~\cite{webster_low_2013} is a recent testament to this. Historically, most attention has been devoted to the development of single-mode ICLs due to their spectral selectivity for unambiguous species identification. However, the limited tuning capability of these single-mode devices has often precluded multi species and broadband sensing, which have become increasingly important in contemporary security, medical diagnostics, and environmental monitoring applications.
     
     \begin{figure*}[ht]
          \centering
          \includegraphics[width=0.96\textwidth]{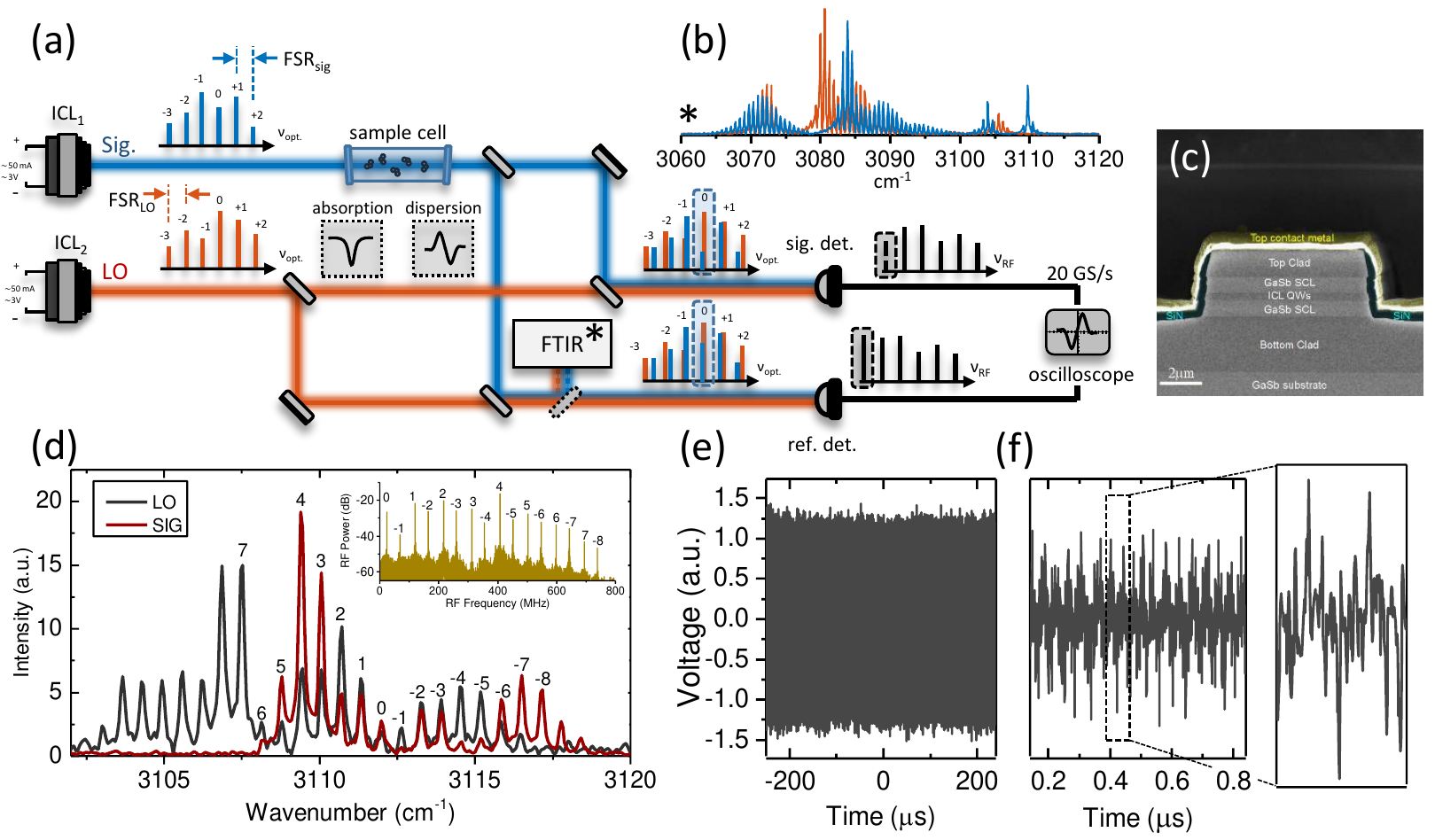}
          \caption{\textbf{Experimental overview and laser characteristics.} \textbf{(a)} schematic of the multiheterodyne setup using interband cascade lasers. One laser is used as a local oscillator, while the other probes the sample. \textbf{(b)} FTIR spectra of the two ICLs driven at temperatures of 17.32$^\circ$C and 11.84$^\circ$C, and currents of 105.3~mA and 127.8~mA, respectively. \textbf{(c)} SEM micrograph of an ICL narrow ridge output facet. \textbf{d,} optical mode spectrum of the ICLs measured by FTIR with a spectral resolution of 0.125~cm$^{-1}$. The intensity of the signal laser is attenuated by propagation through a methane sample. Note that the poor spectral resolution of the FTIR prevents any visual identification of the free spectral range difference of the lasers. The inset shows the resulting multiheterodyne beat spectrum. \textbf{(e)} multiheterodyne interferogram recorded at 20~GS/s~and down-sampled to 2.5~GS/s. \textbf{(f)} expanded view of the data shown in \textbf{(e)}. \label{fig:Fig1}}
     \end{figure*}
 
     An elegant remedy to this issue is provided by the recent advances in multiheterodyne spectroscopy (MHS)~\cite{villares_dual-comb_2014,yang_terahertz_2016} and dual-comb spectroscopy (DCS)~\cite{coddington_dual-comb_2016}, which provide simple and effective means for down-converting spectroscopic information at optical frequencies to the radio-frequency (RF) domain where reliable and convenient measurement instrumentation is readily available. This has long been considered a future alternative to the well-established Fourier transform spectroscopy technique~\cite{griffiths_fourier_2007}, with the benefit of dramatic reductions in acquisition times~\cite{coddington_dual-comb_2016}. Although initially confined to the near-infrared~\cite{coddington_time-domain_2010, coddington_coherent_2008, zhu_real-time_2013}, MHS systems have lately been reported (or proposed) for most wavelengths from the extreme ultra-violet (XUV)~\cite{jones_xuv_2015} trough the mid-infrared~\cite{villares_dual-comb_2014, baumann2011spectroscopy, yan2016mid, zhang2013mid, zhu2015mid} to the terahertz (THz)~\cite{yang_terahertz_2016, yasui_adaptive_2015}. For mid-infrared in-field applications, semiconductor laser technologies such as quantum cascade lasers are of particular interest due to their ability to integrate the dual laser devices monolithically. QCLs have been shown to exhibit frequency comb characteristics if the group velocity dispersion (GVD) is adequately low~\cite{villares_dispersion_2016}, which can be controlled through careful dispersion engineering of the laser cavity~\cite{villares_dispersion_2016, burghoff_dispersion_2016}. If sufficiently low dispersion is obtained, passive frequency comb generation arises through four-wave mixing due to nonlinearities in the laser gain medium~\cite{hugi_mid-infrared_2012}. QCL-based MHS has been demonstrated for detecting various molecules. Villares et al.~\cite{villares_dual-comb_2014} achieved a fractional absorption precision of $\sim$1\%, with a spectral resolution of 80~MHz when water vapor was targeted around 1422~cm$^{-1}$, with more than 15~cm$^{-1}$ of spectral coverage over $\sim$60 optical modes. Assessments of ammonia~\cite{wang_high-resolution_2014, westberg_mid-infrared_nodate}, isobutane~\cite{westberg_mid-infrared_nodate}, and nitrous oxide~\cite{hangauer_wavelength_2016} have also been reported in the literature. However, with laser voltages of $\sim$10~V and biasing currents often exceeding 500~mA these spectroscopic systems consume far too much power to be considered for sustained portable use. 
     
     As an attractive alternative, this work presents the first demonstration of an electronically driven ICL-based MH spectrometer that covers $\sim$240~GHz ($\sim$8~cm$^{-1}$) around 3.2~$\upmu$m. The system is capable of rapid acquisition with response times down to tens of microseconds and ultimate spectral resolution in the MHz range defined by the laser linewidth. As a proof-of-concept, we demonstrate both rapidly-swept single mode absorption and dispersion spectroscopic assessments of low-pressure ethylene (C$_2$H$_4$), as well as $\upmu$s-time-resolved broadband absorption spectroscopy of methane (CH$_4$). In addition, the feasibility of tuning the laser, by current or temperature, to enhance the spectral resolution is explored with methane at atmospheric pressure. 
      
     \section{Experimental procedures}
     
     \subsection{Experimental setup.} Two Fabry-P\'{e}rot interband cascade lasers (ICLs) with mode spacings of 0.640, and 0.643~cm$^{-1}$ ($\Delta$FSR $\approx$ 96 $\pm$ 3~MHz) were mounted in two separate housings (LDM-487201) to prevent thermal crosstalk and enable independent frequency tuning. The light sources' optical spectra are shown in Fig.~\ref{fig:Fig1}(b). The maximum net spectral bandwidths of $\sim$50~cm$^{-1}$ are covered by more than 80 individual modes. The ICL wafer was grown by molecular beam epitaxy at NRL on an n-GaSb (100) substrate, with a 7-stage carrier-rebalanced design~\cite{vurgaftman_rebalancing_2011} similar to that employed by Canedy et al.~\cite{canedy_pulsed_2014}. The narrow ridges were processed by photolithography and reactive-ion etching using a Cl-based inductively coupled plasma (ICP) that proceeded to a depth below the active core of the device, as shown in the end-view SEM micrograph of Fig.~\ref{fig:Fig1}(c). A Si$_3$N$_4$ layer was deposited by plasma-enhanced chemical vapor deposition, and a top contact window etched back using SF$_6$-based ICP. The ridges were metallized and electro-plated with $\sim$5~$\upmu$m of Au, and then mounted epitaxial-side up without facet coatings. Both devices had ridge widths of $\sim$8~$\upmu$m and nominal cavity lengths of 2~mm. However, the two cavities were intentionally cleaved to lengths differing by $\sim$15~$\upmu$m, in order to slightly offset the individual mode spacings to values of 0.640~cm$^{-1}$ and 0.643~cm${-1}$, respectively (corresponding to a free spectral range difference of 96~$\pm$~3~MHz). The cw threshold drive power for each FP-ICL operating at 25$^\circ$C was $\sim$170~mW, which is lower than for a typical QCL but still far higher than in optimized ICL devices. Since the devices were not intentionally dispersion engineered, the multiheterodyne measurements relied on bias and temperature regimes that featured sufficiently narrow beat note linewidths.
     
     Multimode light produced by each ICL, spanning approximately 15~cm$^{-1}$ centered at 3110~cm$^{-1}$, was collimated by an AR-coated f/1.6 aspheric lens (LightPath model 390037IR-3) and strongly attenuated to minimize the effect of nonlinearities in the photodetectors and suppress optical feedback. The use of a custom-made motorized rotary sample cell with individual reference and sample compartments~\cite{smith_real-time_2013} enabled rapid switching between the zero-gas and sample measurements. This ensured that the noise properties of the reference and sample measurements were highly correlated, which reduced the influence of their common mode noise. The sample interaction pathlengths within the absorption cell and reference cell were 2.2~cm. The beams from the two sources were guided into two fast (\=/3~dB cutoff-frequency of $\sim$1~GHz) TEC-cooled mercury-cadmium telluride photodetectors (Vigo PV-4TE-10.6) to provide sample and reference down-converted spectroscopic signals, which were amplified by 40~dB low noise amplifiers (LNA), with noise figures of 1~dB, and bandwidths of 1~GHz (Pasternack PE15A1012). A 3.5~GHz, four channel, 8-bit oscilloscope with a sampling rate of 20~GS/s (LeCroy WavePro 735Zi) was used in oversampling mode to acquire synchronously triggered two-channel spectroscopic data for 500~$\upmu$s with an effective sampling rate of 2.5~GS/s and a theoretical vertical resolution of 12 bits. 
          
     To minimize the RF interference and ground loops, two low-noise laser current drivers (Wavelength Electronics QCL500) were supplied by a 24~V linear power supply (BK Precision 1672) driven from a double-conversion uninterruptable power supply (APC 1000XL) connected via an AC isolating transformer. High precision TEC temperature controllers (Arroyo Instruments 5305 TECSource) were used to minimize thermal fluctuations to below 10~mK. Figure~\ref{fig:Fig1}(a) shows the experimental setup used for this demonstration, where only the optical path is illustrated for simplicity. The setup is built around the differential DCS configuration~\cite{schiller_spectrometry_2002}, where the two high-bandwidth photodetectors are used to suppress amplitude noise through balanced detection. As in most DCS configurations, one device operates as a signal laser that probes the sample species, whereas the other acts as a local oscillator that did not interact with the sample. Careful spatial overlap of the laser beams, together with the difference in mode spacings produces a multiheterodyne beat signal within the photodetector electrical bandwidth ($\sim$1~GHz). This effectively down-converts the optical spectrum information from the optical domain to the RF domain, where it can be conveniently digitized and processed. The parallel nature of this procedure, which provides access to spectroscopic information from all the optical modes simultaneously, sets it apart from most rivaling techniques that require slow single-mode frequency tuning or optical delay line movements. 
     
     Fig.~\ref{fig:Fig1}(d) shows the emission spectrum of the ICLs measured by Fourier transform spectroscopy with a resolution of 0.125~cm$^{-1}$. The pairs of modes from both lasers creating heterodyne beat notes within the detector bandwidth are labeled with integers for identification, and the corresponding RF beat notes are shown in the inset. The pair of optical modes with the highest spectral overlap (labeled with '0') produce the lowest frequency beat note, which is positioned explicitly at a frequency corresponding to a quarter of the free spectral range difference. Since this mode resides near the center of the mode structure, its precise positioning in the RF domain enables aliasing of the positive and negative frequencies in the DC-centered RF spectrum, which creates equidistant frequency spacing of the beat notes. A doubling of the number of beat notes available within the detector bandwidth is thereby achieved, which subsequently doubles the optical spectral coverage~\cite{schiller_spectrometry_2002}. For the particular laser operating conditions used, 16 beat notes could be acquired simultaneously, which provides an instantaneous spectral coverage of $\sim$300~GHz. Fig.~\ref{fig:Fig1}(e) shows a time-domain interferogram acquired during 500~$\upmu$s. Just as in QCL-based dual comb spectroscopy, the amplitude is nearly constant during the acquisition~\cite{villares_dual-comb_2014}, which is in significant contrast to MHS systems based on mode-locked dual comb lasers, where a burst is observed at zero delay time~\cite{coddington_coherent_2010}. Fig.~\ref{fig:Fig1}(f) shows an expanded view of the interferogram, where the repetitive character of the MH beating signal is visible.

          \begin{figure*}[h!t]
          	\centering
          	\includegraphics[width=1\textwidth]{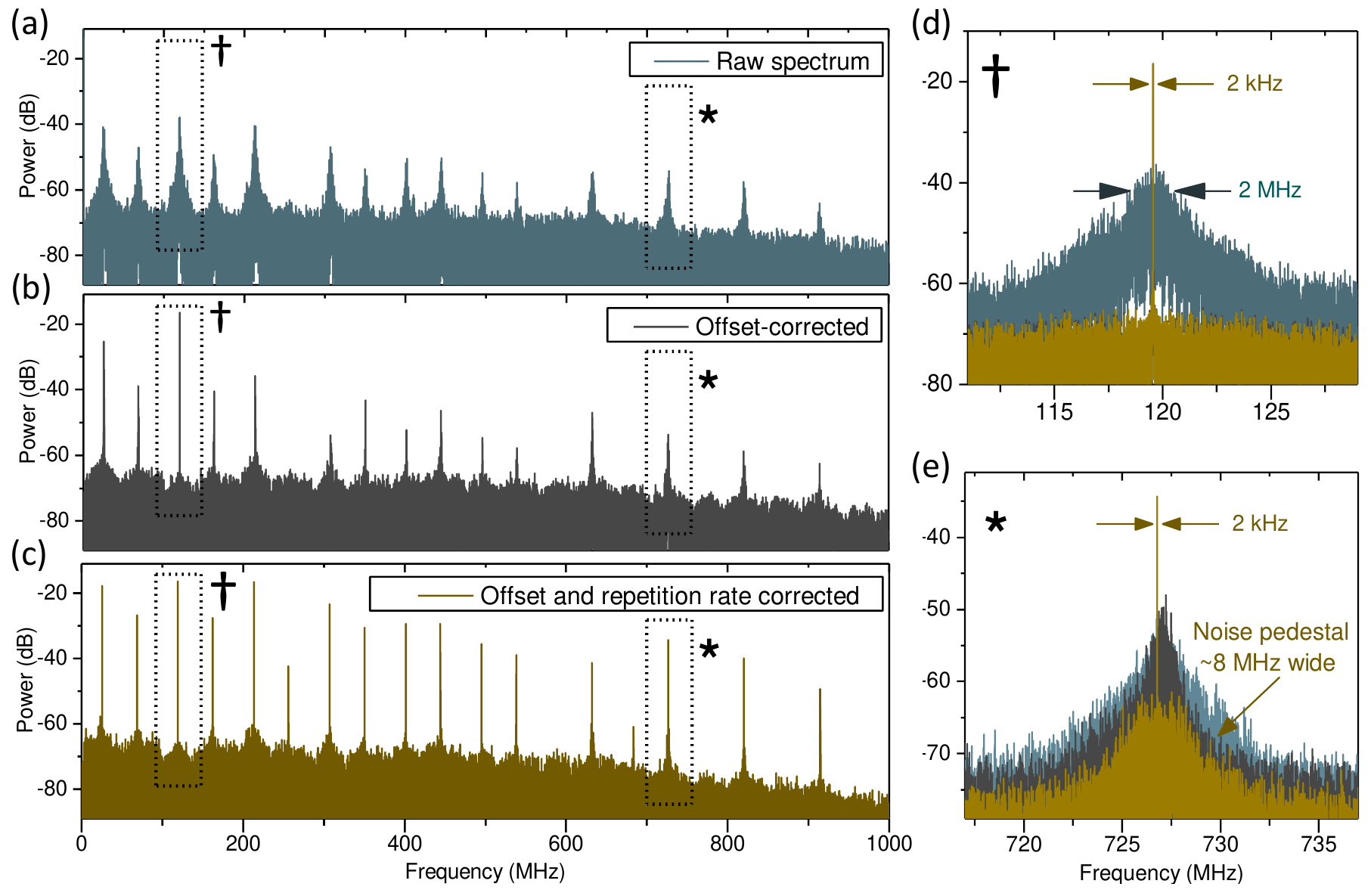}
          	\caption{\textbf{Computational phase and timing corrections.} \textbf{(a)} multiheterodyne beat note spectrum without any corrections. The phase noise pedestal is clearly observable, and the beat note amplitude SNR is $\sim$20~dB. \textbf{(b)} multiheterodyne beat note spectrum after frequency offset correction. Most of the phase noise is reduced by this procedure. \textbf{(c)} multiheterodyne beat note spectrum after correcting both the offset and repetition rate. Most of the remaining phase noise is canceled and the signal-to-noise ratio is visibly increased, which is especially clear for low amplitude beat notes. \textbf{(d)} expanded view of the beat note located at 120~MHz. The beat note linewidth is reduced by three orders of magnitude, from 2~MHz to 2~kHz (lower-bound by the Fourier limit). The beat note SNR ratio exceeds 50~dB for the corrected case, compared to $\sim$20~dB for the uncorrected case. \textbf{(e)} expanded view of the beat note at 726~MHz. Due to the greater frequency spacing from the beat notes used for correction, the phase noise cancellation is less effective. However, the beat note SNR still exceeds 30~dB. \label{fig:Fig2}}
          \end{figure*}     
      
     \subsection{High resolution narrowband direct absorption and dispersion spectroscopy.} The high resolution capabilities were evaluated by extracting single beat note Doppler limited direct absorption and dispersion spectra, where the two lasers were frequency tuned by collinearly ramping the injection currents at a rate of 1~kHz, which yielded a symmetrical scan of $\sim$1~GHz around the center frequency of the targeted low pressure (1.6~kPa) ethylene transition at 3112.58~cm$^{-1}$. It should be noted that such high-resolution spectrum can in principle be acquired simultaneously with all available beat notes, but the acquisition equipment available for this demonstration allowed only single beat note detection. The signal acquisition was performed with an 8-bit oscilloscope (LeCroy WavePro 735Zi) and the acquisition time was set to 1~ms to capture the rising and falling edges of the scan. For these measurements a 100~mm long sample cell (PIKE technologies) was used to allow sufficiently high absorptions at low pressures (Doppler limited regime). For both dispersion and absorption detection, the reference detector channel was used as an optical power reference. For the absorption spectra, a sample gas spectrum and a zero gas (nitrogen) were acquired consecutively to enable baseline correction. In the case of the dispersion spectra, an adjacent (unaffected) beat note was instead used as a phase reference. The phase and amplitude information were obtained by computationally IQ-demodulating the pertinent beat notes within a bandwidth of 10~MHz around the center beat note frequency.     
      
     \subsection{HITRAN Simulation.} The spectra of methane and ethylene were obtained based on simulations using spectroscopically relevant parameters from the HITRAN 2012 database~\cite{rothman_hitran2012_2013}. The simulations assumed a temperature of 296~K and partial pressures of 99~kPa and 1.6~kPa for methane and ethylene, respectively. A Voigt lineshape was assumed and no apodization function was used.
      
     \subsection{Computational phase and timing correction.} To perform reliable and sensitive multiheterodyne measurements, it is imperative to achieve sufficient phase-noise coherence between the two interacting light sources. A failure to meet this requirement significantly broadens the multiheterodyne beat notes, which may preclude sensitive spectroscopic assessments. Ideally, a stable phase reference should be used to actively control the phase and timing errors of the two combs. However, such a procedure is experimentally cumbersome, and may not yield an improvement sufficient to justify its complexity. Instead, purely computational methods can be implemented to coherently correct the multiheterodyne spectrum, as recently demonstrated by Burghoff et al.~\cite{burghoff_computational_2016}. While this approach requires no a priori information regarding the combs, its Kalman filter approach can be computationally demanding when handling a large number of beat notes. Here, we developed a phase and timing correction methodology that is less computationally intense and can be used reliably in cases where the phase and timing errors are relatively small (more details of the phase and timing correction algorithm are provided in the supplementary information). The core of this method is the assumption that the multiheterodyne spectrum exhibits frequency comb characteristics, i.e., it can be uniquely described by two frequencies: the relative offset frequency ($\Delta f_{ceo}$) and the relative repetition rate frequency ($\Delta f_{rep}$). The phase and timing errors can then be extracted and corrected by considering the instantaneous phases of only two adjacent beat notes, whose phases and phase difference are used for the global phase correction (relative offset frequency correction), and for the timing correction (i.e. repetition rate correction, which is similar to adaptive sampling~\cite{yasui_adaptive_2015,ideguchi_adaptive_2014}), respectively. 
     
     Results from each correction step are displayed in Fig.~\ref{fig:Fig2}, where (a) shows the uncorrected multiheterodyne spectrum obtained by Fourier transforming the time-domain signal of Fig.~\ref{fig:Fig1}(e). Although a large portion of the phase noise can be suppressed solely by correcting the offset frequency, some remnants due to the repetition rate fluctuations can clearly be observed for beat notes at higher RF frequencies. This effect is clearly apparent in Fig.~\ref{fig:Fig2}(b), for which only the relative offset frequency correction is applied. Further improvement is possible by correcting the repetition rate fluctuations. The RF spectrum obtained after this operation is shown in Fig.~\ref{fig:Fig2}(c). Naturally, the largest linewidth reductions are obtained in proximity to the two beat notes used for the correction, as shown in Fig.~\ref{fig:Fig2}(d) where the initial linewidth of 2~MHz is reduced to~2 kHz (limited by the resolution bandwidth of the acquisition). For beat notes at higher frequencies the -3~dB linewidth reduction is similar, but a phase noise pedestal $\sim$10~dB above the noise floor remains after the correction [see Fig.~\ref{fig:Fig2}(e)]. The less efficient phase noise cancellation far from the correction pair can be attributed to non-ideal retrieval of the repetition rate signal and accumulative error propagation, or lower phase-noise coherence of the beat notes, which may be addressed by optimization of the laser cavity dispersion~\cite{villares_dispersion_2016, burghoff_terahertz_2014, burghoff_dispersion_2016}. The latter phenomenon has been observed also for the more computationally intensive algorithm of Burghoff/Yang et al.~\cite{burghoff_computational_2016}. Despite the remaining phase noise pedestal, an enhancement of signal-to-noise ratio (SNR) by up to $\sim$30~dB and beat note linewidths down to kHz level are attainable using the method proposed here, which is sufficient for most spectroscopic applications. Also, the narrow beat note linewidths indicate that FP-ICLs exhibit a similar optical phase-locking mechanism as observed in QCLs~\cite{hugi_mid-infrared_2012}. A detailed description of the phase and timing correction procedure can be found in the Appendix.
          
     \subsection{RF beat note lock.} Owing to a large drift of the offset frequency throughout the measurement, an active beat note locking circuitry was implemented to keep the RF spectrum at a constant frequency position~\cite{sterczewski_molecular_2017}. A bandpass-filtered beat note located at a quarter of the FSR difference was fed to a custom-made broadband frequency discriminator utilizing a first-order LC filter, followed by a logarithmic gain detector (Analog Devices, AD8302) outputting a frequency-dependent voltage suitable for an analog high-bandwidth (100~kHz) PID controller (SRS SIM960). The output from the controller, after 20~dB attenuation, was used as an injection current modulation signal to the local oscillator (LO) laser driver to compensate for the frequency drifts. This locking procedure was implemented for all spectroscopic measurements. Note that this only stabilizes the beat notes to fixed frequencies, but does not alter their linewidths, in contrast to high-bandwidth phase-locking~\cite{westberg_mid-infrared_nodate}, where narrowing of the linewidths also occurs. 
          
     \section{Results}     
     
     \subsection{Beat note frequency and amplitude stability.} 
\begin{figure}[H]
     	\includegraphics[width=.49\textwidth]{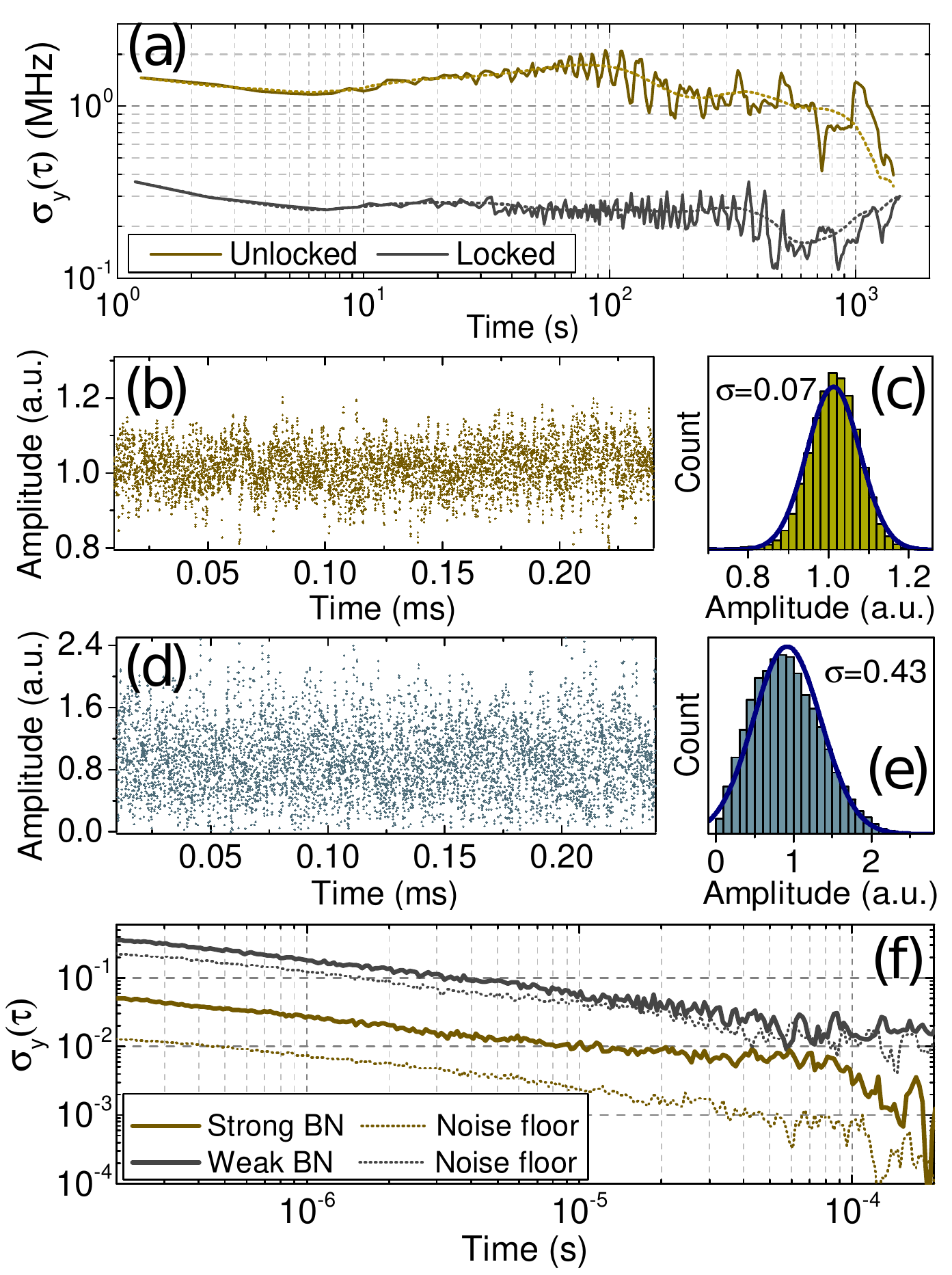}
     	\caption{\textbf{Beat note frequency and amplitude stability.} \textbf{(a)} Allan deviation of the beat note center frequency. A noticeable improvement is observed by enabling the frequency discriminator locking procedure. \textbf{(b)} amplitude measurements obtained through computational IQ-demodulation of a comparatively strong beat note (at 120~MHz in Fig.~\ref{fig:Fig2}). \textbf{(c)} the distribution of the beat note amplitudes displayed in \textbf{(b)}. The distribution is of normal type, which is typically encountered for high-SNR beat notes. \textbf{(d)} amplitude measurements of a comparatively weak beat note (at 257~MHz in Fig.~\ref{fig:Fig2}). \textbf{(e)} the distribution of the beat note amplitudes displayed in \textbf{(d)}. Due to rectification, the distribution is skewed, which is common for weaker beat notes. \textbf{(f)} The solid lines represent typical relative beat note amplitude stabilities for weak and strong beat notes, whereas the dashed lines represent the relative amplitude stability of the adjacent noise floor (obtained through IQ-demodulation detuned from the beat note). \label{fig:Fig3}}
     \end{figure}

Long-term frequency drifts of the beat notes' center frequencies may significantly deteriorate the reliability of the optical mode assignment, due to the beat note aliasing commonly used in DCS~\cite{schiller_spectrometry_2002}. Drifts of a quarter of the free spectral range difference between the two lasers will effectively cause an overlap of the multiheterodyne beat notes, which hinders individual amplitude and phase extraction. Due to the finite linewidth of individual beat notes, a conservative maximum frequency drift of 20~MHz is desirable in this case. The beat note frequency stabilization was accomplished through frequency stabilization of the LO laser with respect to the signal laser. While operating the lasers in relative frequency locked-mode, the beat note center frequency stability was evaluated using Allan deviation measurements~\cite{werle_limits_1993}, which are shown in Fig.~\ref{fig:Fig3}(a). The system exhibits a beat note center frequency drift below $\sim$300~kHz in locked-mode, and the long-term drifts are far below the aforementioned 20~MHz. 
      
     An important evaluation metric for any multiheterodyne spectrometer is the beat note amplitude stability~\cite{newbury_sensitivity_2010}, which limits the sensitivity under normal operation. In order to assess this, Allan deviation analysis was performed for all the multiheterodyne beat notes of Fig.~\ref{fig:Fig2} by IQ-demodulating the instantaneous amplitude of the phase-corrected signal within a 3~MHz bandwidth. The Allan deviation of the strongest and weakest beat notes are shown in Fig.~\ref{fig:Fig3}(f), and the corresponding amplitude data as a function of acquisition time are shown in Fig.~\ref{fig:Fig3}(b) and (d), respectively. Fig.~\ref{fig:Fig3}(c) and (e) display the histograms of the beat note amplitudes, where the low-SNR beat note clearly displays nonconformity to the normal distribution. This implies that in general a simple mean evaluation of the beat notes should be avoided. 
     
     The Allan deviations indicate white-noise limited performance of acquisitions up to 100~$\upmu$s, which yields percentage level precision even for the low-SNR beat notes. The Allan analysis can also be used to estimate the system performance for a given acquisition time, which determines the ultimate time resolution of the system for a given detection limit. In this case, according to the stability analysis, the weak beat notes suffer a precision reduction from $\sim$2\% at 100~$\upmu$s to $\sim$5\% at 20~$\upmu$s (or a lower bound of a bandwidth-normalized noise-equivalent absorption NEA $\approx$ 2$\times$10$^{-4}$/Hz$^{1/2}$), whereas corresponding values for the strong beat note are from $\sim$0.5\% at 100~$\upmu$s to $\sim$1\% at 20~$\upmu$s (or NEA $\approx$ 5$\times$10$^{-5}$/Hz$^{1/2}$). The values are comparable to earlier reports on QCL-based MHS~\cite{villares_dual-comb_2014, westberg_mid-infrared_nodate, hangauer_wavelength_2016, sterczewski_molecular_2017} and mid-infrared microresonator DCS~\cite{yu2016silicon}.

     \begin{figure*}[h!t]
     	\centering
     	\includegraphics[width=1\textwidth]{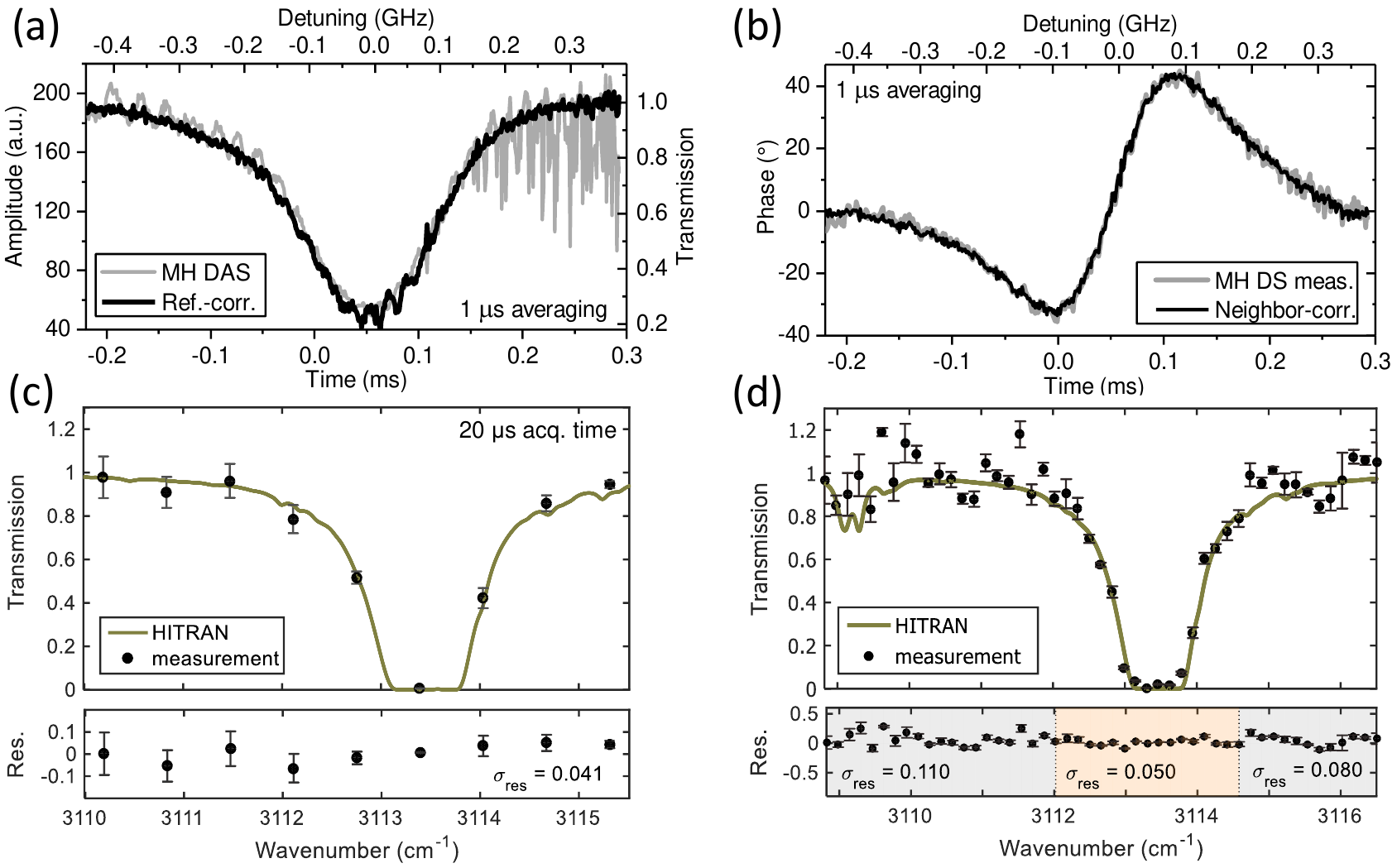}
     	\caption{\textbf{Multiheterodyne spectroscopy.} (\textbf{a}) swept direct absorption multiheterodyne spectroscopy of 1.6~kPa ethylene at 3112.6~cm$^{-1}$. An off-resonance noise estimation yields a SNR of $\sim$18~dB. The measurement was acquired using a 1~kHz triangular ramp, which gives an acquisition time of 500~$\upmu$s per spectrum. (\textbf{b}) swept dispersion multiheterodyne spectroscopy using the same time-domain data as in a. An adjacent beat note unaffected by the molecule was used as a phase reference. The off-resonance noise estimation gives a SNR of $\sim$20~dB. (\textbf{c}) broadband multiheterodyne spectroscopy of 99~kPa methane using an acquisition time of 20~$\upmu$s shown with HITRAN model residuals. (\textbf{d}) broadband spectrum of methane achieved by temperature tuning to interleave the individual measurements for finer spectral resolution. The standard deviation of the fit residuals directly reflects the signal-to-noise ratio of the beat notes, resulting in lower standard deviations near the center of the spectrum. A detailed derivation of 1$\sigma$ confidence intervals for the data points in (\textbf{c}) and (\textbf{d}) is described in Ref. ~\cite{westberg_mid-infrared_nodate} \label{fig:Fig4}}
     \end{figure*}
     
     \subsection{Narrowband multiheterodyne spectroscopy.} The versatility of the mid-IR FP-ICL based MHS technique is demonstrated by performing single-mode acquisitions, analogous to the well-developed tunable diode laser absorption spectroscopy (TDLAS) techniques~\cite{fried_infrared_2006}. In conventional TDLAS, the frequency of the probing laser is swept across an absorption feature, which encodes the spectroscopic information in the attenuated intensity. Similarly, in swept direct absorption MHS, both lasers are rapidly swept across a narrowband transition (<1~GHz) while the amplitude of their resulting beat note is analyzed. Since this can be accomplished without any opto-mechanical alteration to the system, the broadband and narrowband measurements can be toggled seamlessly. For demonstration, low pressure ($P$=1.6~kPa, $T$=21$^\circ$C) ethylene (C$_2$H$_4$) was used as the pilot species and the lasers were collinearly frequency scanned in locked mode at a rate of 1~kHz. The spectroscopic information was deduced from the time-domain signal acquired by a 20~GS/s oscilloscope. Computational IQ-demodulation with a bandwidth of 3~MHz around the attenuated beat note frequency gives direct access to the amplitude and phase interactions of the light with the sample. The results of this procedure are given in Figs.~\ref{fig:Fig4}(a) and (b), where (a) shows the amplitude response re-calculated as transmission while (b) shows the phase change induced by the interaction with the absorber. It should be noted that the phase information was extracted by using an unaffected adjacent beat note as a reference. This type of differential measurement effectively suppresses any common mode noise that appears in the phase response, thereby reducing the influence of transmission fluctuations~\cite{sterczewski_molecular_2017}. The spectrometer achieves a bandwidth-normalized noise-equivalent absorption of 3.3$\times$10$^{-4}$/Hz$^{1/2}$) when operating in swept absorption mode, and 2.3$\times$10$^{-4}$/Hz$^{1/2}$) in dispersion mode using scan rates of up to 1~kHz.  

     \subsection{Rapid broadband multiheterodyne spectroscopy.} The beat note stability analysis described earlier indicates that $\upmu$s time-resolution is feasible with the ICL-based multiheterodyne spectrometer. To validate this, a 20~$\upmu$s measurement of methane was performed under similar conditions to those described in the previous Section. The results are displayed in Fig.~\ref{fig:Fig4}(c), where the residuals in the lower panel indicate a NEA of 2$\times$10$^{-4}$/Hz$^{1/2}$). Although an initial reference measurement was used for calibration, this experiment nonetheless demonstrates the feasibility of rapidly acquiring data from transient chemical reactions such as those encountered in combustion processes.

     \subsection{Broadband multiheterodyne spectroscopy.} The feasibility of high resolution broadband gas sensing by performing interleaved spectral measurements of a methane (CH$_4$) sample at 99~kPa ($T$=21$^\circ$C) has been investigated. The sample was placed in a custom-made motorized rotary cell with separated reference and sample compartments~\cite{smith_real-time_2013}. This allowed for rapid subsequent zero-gas reference measurements using ambient air as the reference gas ($P_{tot}$=101.3~kPa, $T$=21$^\circ$C). In addition to normalizing the power, the reference detector signal also provided access to the beat note used to stabilize the relative frequency of the lasers~\cite{sterczewski_molecular_2017}. 
     
     In contrast to single-mode swept MHS, where the spectral resolution is limited by the linewidths of the optical modes, the spectral sampling resolution of a broadband multiheterodyne spectrometer can be no finer than the laser mode spacing, in our case $\sim$0.64~cm$^{-1}$ (19.2~GHz). However, this can be mitigated by using the ICL's frequency tuning properties (either through current or temperature) to interleave several spectra over the entire free spectral range (FSR)~\cite{villares_dual-comb_2014}; although changes in the laser cavity dispersion characteristics~\cite{villares_dispersion_2016} due to tuning may affect the performance. As a proof-of-concept, a step-scan (one quarter of the FSR) was performed through tuning the laser temperatures, thereby achieving a higher spectral sampling resolution assessment of the methane absorption. The results shown in Fig.~\ref{fig:Fig4}(d) indicate good agreement with the HITRAN~\cite{rothman_hitran2012_2013} simulated spectrum for the strong beat notes around the center, where remaining discrepancies are primarily due to optical fringes in the sample cell windows and low beat note SNRs for some of the temperatures. 
     
     \section{Discussion}
     A compact, energy-efficient, solid-state multiheterodyne spectrometer based on multimode FP-ICLs has been demonstrated. The performance of the spectrometer has been evaluated by measuring the broadband spectrum of methane at atmospheric pressure with an instantaneous optical bandwidth of ~240 GHz. In addition, the fully electronically controlled system can be seamlessly altered for narrowband detection with similar performance as systems using conventional single-mode lasers. This capability is investigated by measuring the narrowband absorption and dispersion of low pressure ethylene. In this approach, the inherent common-mode noise rejection of the differential dispersion measurement shows great potential for applications limited by transmission fluctuations. These noise sources are typically abundant in environments heavily contaminated by particulate matter. Both swept and rapid broadband modes of operation consistently exhibited NEAs of $\sim$3$\times$10$^{-4}$/Hz$^{1/2}$, which is close to the amplitude noise limit given by the fluctuations of the beat notes' amplitudes (see Fig.~\ref{fig:Fig3}). 
     
     Since the FP-ICLs exhibit similar phase locking mechanism as earlier reported QCL combs~\cite{hugi_mid-infrared_2012}, it is expected that via careful dispersion engineering of the ICL gain and optimization of laser cavities, similar broadband spectral coverage and low noise operation can be achieved, consequently improving the MHS sensitivity. This, in combination with the low power consumption of the ICLs, will undoubtedly prove useful for applications where available power is scarce. A typical scenario would be the nodes in autonomous sensor networks, or portable sensors that require battery operation or solar power.

	\section*{Appendix: Computational phase and timing correction in multiheterodyne spectroscopy}
	\label{sec:appendix}
     
	 Several different phase correction methods have been suggested in order to perform coherent averaging~\cite{coddington_dual-comb_2016,ideguchi_adaptive_2014,rompelman1986coherent,hebert2017self} in multiheterodyne (MH) systems by suppressing the unwanted measurement instabilities of free running frequency combs. Recently, Burghoff/Yang et al.~\cite{burghoff_computational_2016} proposed a computational phase and timing correction methodology based on Kalman filtering, which can handle extreme multiheterodyne beat note instabilities and RF spectrum congestion due to overlapping beat notes. Despite the excellent performance of this method there are some practical limitations due to the computational complexity of the Kalman filter, as well as the requirement of a priori knowledge of the number of modes, $N$, contributing to the RF spectrum, which may limit its real-time phase correction performance even in systems with a lower number of multiheterodyne beat notes.
	  
	 Under the condition that the observed RF spectrum contains resolvable beat notes (which may be affected by a significant amount of phase noise), a simpler computational procedure can be implemented to perform the phase correction. Similarly to the methodology by Burghoff/Yang et al. this technique also leads to an improvement of the beat note signal-to-noise ratio and provides compensation for multi-peak multiheterodyne beat notes produced by sudden changes in the repetition frequency~\cite{burghoff_coherent_2015}. The major advantages of the proposed method are low computational complexity, feasibility of a pure hardware on-line implementation, and no need for a priori knowledge about the instability characteristics.
	
     Since the multiheterodyne beat notes may suffer from large frequency offset fluctuations, which broaden the beat notes significantly up to MHz linewidths (tens of MHz for the phase noise pedestal), the spacing between the beat notes in the RF domain needs to be sufficient to resolve adjacent frequency components. In order to fully utilize the bandwidth of the photodetectors, the spectral folding scheme proposed by Schiller~\cite{schiller_spectrometry_2002} can still be employed. This aliases the beat notes such that they remain uniformly spaced, which is commonly accomplished by DC-centering the RF beat notes and tuning the lowest-frequency beat note to a frequency corresponding to a quarter of the free spectral range difference ($\Delta$FSR or repetition rate $\Delta f_\mathit{{rep}}$). Preferably, the RF beat notes should be loosely frequency-locked to this position, which greatly enhances the long-term stability of the digital correction algorithm. The spectral folding procedure allows for a doubling of the number of usable beat notes within the detection bandwidth, at the expense of slightly complicating the signal processing.
     
	 \subsection*{A.1 Correction algorithm}
     
     To correct for the relative phase fluctuations of the multimode sources, the algorithm assumes a frequency comb model. The instantaneous frequency of the n$^{\mathrm{th}}$ comb line can be expressed as
     \begin{equation}
     f_n = f_0 + nf\;,
     \end{equation}
     where $n$ is a large integer, $f$ is the mode spacing, and $f_0$ is the offset frequency. As a result of the optical beating between the two sources, the optical spectrum is down-converted to the RF domain, producing a comb of frequencies spaced by $\Delta f_\mathit{{rep}} = |f_\mathit{{rep1}}-f_\mathit{{rep2}}|$, which is the difference in the free spectral ranges ($f_\mathit{{rep1}}$ and $f_\mathit{{rep2}}$) of the two laser sources used in the system). Therefore the instantaneous RF frequency of the n$^{\mathrm{th}}$ beat note can be expressed as:
	 \begin{equation}
	 f_{n,rf} = f_\mathit{{off}} + n\Delta f_\mathit{{rep}}\;.
	 \end{equation}
     In the RF domain, all the beat notes are shifted equally in frequency by the difference in offset frequencies of the individual lasers $f_\mathit{{off}} = |f_\mathit{{0,A}}-f_\mathit{{0,B}}|$, whereas the fluctuations of the particular beat note frequency due to variations in $\Delta f_{rep}$ are dependent on the beat note number $n$. This effectively causes variations in the length of the interferogram and requires correction of the time axis to enable coherent averaging of the spectra~\cite{ideguchi_adaptive_2014, rompelman1986coherent},. 
    
          \begin{figure}[H]
          	\includegraphics[width=.49\textwidth]{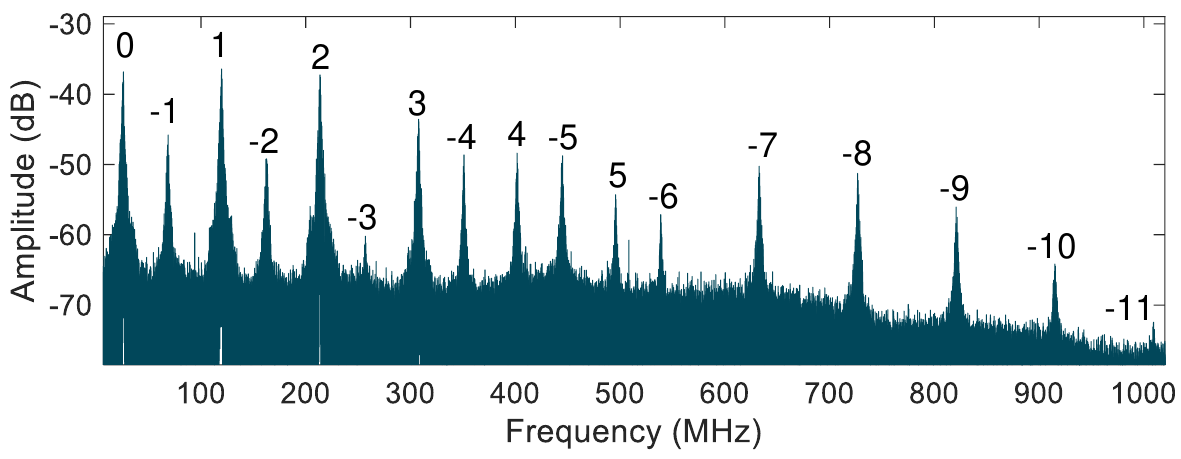}
          	\caption{\textbf{Negative beat notes are folded} to positive frequencies, which can easily be verified through their reversed frequency tuning.  \label{fig:Fig5}}
          \end{figure}          
         
     The bandwidth-maximizing spectral folding scheme introduces an additional complexity since any frequency fluctuations in the folded frequency components exhibit the reversed frequency dependence compared to the non-folded ones. This originates from the non-negative frequency representation (folding), \textit{i.e.},     
	 \begin{equation}
	 f_\mathit{{n,rf}} = |f_{\mathit{off}} + n\Delta f_{rep}|\;.
	 \end{equation}
	 where $n$~=~($\ldots$ , -2, -1, 0, 1, 2, $\ldots$), and the negative signs correspond to the folded frequency components. Consequently, the phase fluctuations of the folded and non-folded beat notes have opposite phase and need to be corrected separately. The correction procedure is performed in two stages: first the repetition rate fluctuations are corrected, which removes the dependence on $n$, after which the relative offset frequency fluctuation correction can be performed using any beat note. In the proposed phase and timing correction scheme, instead of utilizing the time-domain signal resulting from a superposition of all modes, the correction signals are extracted by digital mixing and filtering applied to the two beat notes with the highest signal-to-noise ratio (beat notes '1' and '2' in Fig.~\ref{fig:Fig5} are used in this example).     
	 	 
          \begin{figure}[H]
          	\includegraphics[width=.49\textwidth]{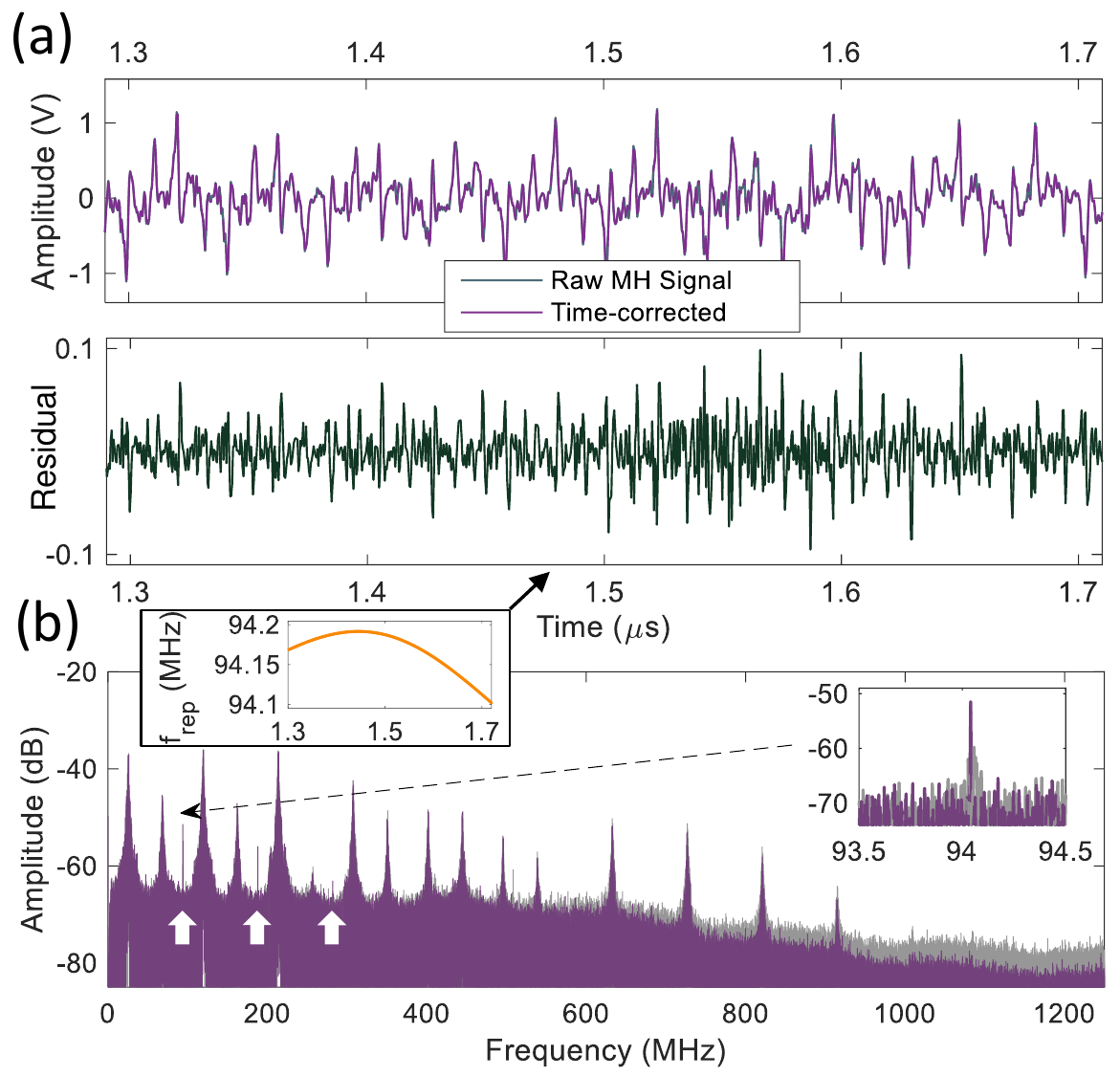}
          	\caption{\textbf{Timing correction of the interferogram.} (\textbf{a}) Zoom of the interferogram before and after the correction. The repetition rate in the presented time window is slightly greater than the mean 94.04~MHz, therefore the resampling procedure stretches the signal in time. The difference between the original and resampled signal is shown in the residual plot, where the cumulative effect of a locally increased speed causes an increasing difference between the two channels. This gradually decreases as the repetition frequency tends towards the mean. (\textbf{b}) In the spectral domain, the detector signal possesses clear harmonics of the repetition frequency as a result of self-mixing of the RF beat notes. For visibility, the self-mixing beat notes are marked with white arrows together with an inset showing the narrowing of the self-mixing beat note. This is an indication that the beat notes jitter coherently and only the offset frequency fluctuations remain to be corrected. \label{fig:Fig6}}
          \end{figure}	 	 
          
     \subsection*{A.2 Timing correction}
	 
	 The repetition frequency correction is performed by timing correction. The general principle of this process could be compared to an analog magnetic tape, containing a recording of an amplitude-modulated single frequency carrier tone. If, due to aging, the tape material has been locally stretched or compressed, the frequency of the physically recoverable carrier tone will be affected. To correct for these fluctuations during the recovery of the original amplitude message, one needs to play the tape at a variable rate, which assures constant carrier frequency. If an average frequency is assumed or known \textit{a priori}, or is assumed to be the average of that on the tape, one can play the recording at a rate $r$ given by 
	 \begin{equation}
	 r(t) = \frac{\langle f_\mathit{carrier}\rangle}{f_\mathit{carrier}(t)}\;.
	 \end{equation}
	 where $f_\mathit{carrier}$ denotes the instantaneous carrier frequency, and $\langle f_\mathit{carrier}\rangle$ is its expected value. In our experiment, similar correction can be applied to correct for fluctuations in $\Delta f_\mathit{rep}$, where the variable rate $r$ is analogous to a nonlinear time axis $t'$, where the time flow is perturbed according to 
	 \begin{equation}
	 t'(t)=\int_0^t \frac{\langle\Delta f_\mathit{rep}\rangle}{f_\mathit{rep}(\tau)} d\tau   \;.
	 \end{equation}
	 Such a formulation has a straightforward application in the discrete data case, where the signal is sampled at a constant time interval T, which requires resampling at new time points given by $t'$,
	 \begin{equation}
	 t'(nT)=T\cdot \sum_{k=1}^n \frac{\langle\Delta f_\mathit{rep}\rangle}{f_\mathit{rep}(kT)} \;.
	 \end{equation}
	 This adaptive sampling scheme can be realized on-line by triggering the multiheterodyne signal acquisition at $\Delta f_\mathit{rep}$~\cite{ideguchi_adaptive_2014, ideguchi2012adaptive}, or by linear resampling in post-processing. The latter method has been used here to eliminate the need for any additional optical or electronic elements in the spectroscopic system.  
	 	 
	 To extract the $\Delta f_\mathit{rep}$ fluctuations in a computational manner, a peak detection procedure is implemented to determine the center frequencies of the two dominant RF beat notes (spaced by the mean value of $\Delta f_\mathit{rep}$). This is followed by bandpass filtering at the extracted center frequencies (filter bandwidth set to 80\% of $\Delta f_\mathit{rep}$/2, here $\sim$37.6~MHz) via a digital high-order zero-phase Butterworth IIR filter that processes the signal in both the forward and reverse direction. This ensures a linear phase response of the recursive filter~\cite{gustafsson1996determining} and removes unwanted time-shifts introduced by the filter. The IIR structure makes it possible to lower the filter order compared to a conventional FIR structure with the same stopband attenuation. In addition, it minimizes the computational cost of the filtering procedure, albeit at the expense of edge effects that can be suppressed by appropriate padding of the signal. The filtered beat note signals are mixed by squaring the amplitude versus time signal and passing the differential frequency component only. Finally, a Hilbert transform is applied to the mixed time-domain signal to create its complex representation for the instantaneous phase retrieval. 
	 
	 The instantaneous repetition frequency is calculated from the instantaneous phase by a Savitzky-Golay derivative, whose filter order is adjusted so that it rejects numerical artifacts caused by differentiation without over-smoothing the estimate. Figure~\ref{fig:Fig6} shows the influence of the timing correction on the interferogram signal and its RF spectrum. Due to residual nonlinearities in the RF signal path the high phase correlation of the multiheterodyne beat notes results in a detectable self-mixing signal at $\Delta f_\mathit{rep}$. After the timing correction, the self-mixing signal is clearly amplified and narrowed down to widths ultimately limited by the acquisition time, which also results in enhancement of its higher harmonics that were barely visible before the correction (see the inset in Fig.~\ref{fig:Fig6}b).

	 To further improve the performance of the timing correction, the dominant mode pair spacing criterion can be extended to multiples of $\Delta f_\mathit{rep}$. This also allows for the use of non-neighboring RF beat notes since the high SNR beat notes may lie in different parts of the spectrum. However, in most cases the neighboring RF beat notes are of comparable amplitude, making the simplified single $\Delta f_\mathit{rep}$ spacing case sufficient.  
	
          \begin{figure*}[h!tb]
          	\centering
          	\includegraphics[width=1\textwidth]{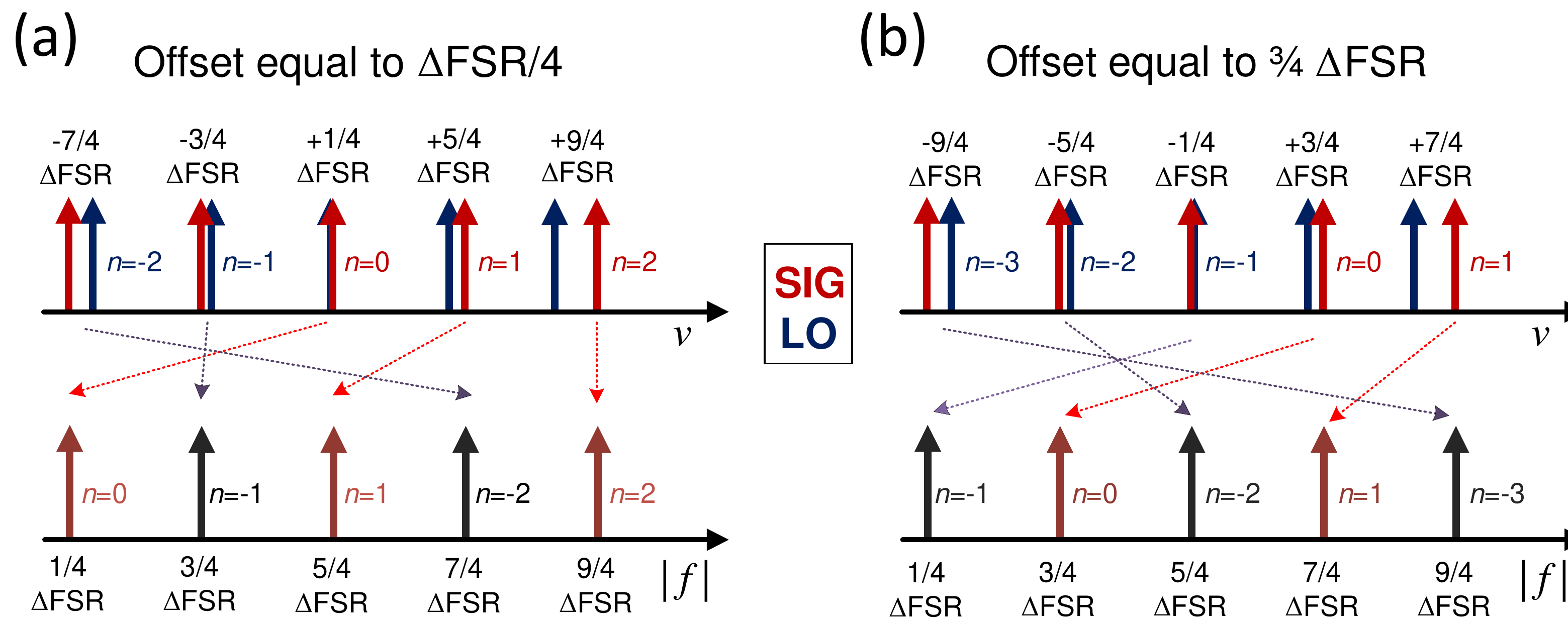}
          	\caption{\textbf{Ambiguity of the RF-to-optical domain mapping order in the spectral folding scheme.} (\textbf{a}) the lowest-frequency beat note is a result of placing the signal comb at a higher frequency than that of the LO. In this case, the RF-to-optical domain mapping is consistent with the well-established folding scheme. (\textbf{b}) the lowest-frequency beat note results from placing the signal comb at a lower frequency compared to the LO. In this case, the lowest frequency beat note will map a folded mode. The two cases look identical in the RF domain, but they map the optical domain differently. This can be distinguished and controlled by a frequency discriminator locking scheme (see text). \label{fig:Fig7}}
          \end{figure*}   	 	
	 
	 \subsection*{A.3 Spectral unfolding}
	 
	 If the multiheterodyne system uses the spectral folding technique, spectral unfolding is required in order to correct the phases of both folded and non-folded beat notes in a single step. If the RF beat notes are loosely frequency stabilized, the beat note spacing and individual frequencies are well-defined, which simplifies the unfolding procedure (first beat note frequency and the repetition rate define all frequencies in the spectrum). This approach was used in this work.
	 There are two possible configurations of the folded multiheterodyne RF spectrum, which differ in the order the beat notes map the optical domain to the RF domain. Subsequently, these configurations determine which direction the beat notes move when the offset frequency is changed, as shown in Fig.~\ref{fig:Fig7}. The most widely used convention, proposed by Schiller~\cite{schiller_spectrometry_2002}, places the optical comb interacting with the sample (whose mode spacing is greater than that of the LO) so that the optical frequency of the mode that produces the lowest frequency beat note is located at $\Delta$FSR/4 higher in frequency relative to the LO (see Fig.~\ref{fig:Fig7}(a)). Thereby the optical domain is mapped to the RF domain from low to high frequency, \textit{i.e.}, the optical modes are numbered from negative to positive in the order they appear in the spectrum. In this configuration, the jitter of the lowest frequency beat note becomes just the relative offset frequency fluctuation $f_\mathit{off}$), and all the related RF beat notes spaced by $\Delta f_\mathit{rep}$) will map the positive-numbered optical modes in a non-mirrored fashion. However, one can easily obtain the opposite situation (Fig.~\ref{fig:Fig7}(b)), where instead the aforementioned optical mode is lower in frequency than the corresponding LO mode. The resulting frequency of the lowest-frequency RF beat note will thus be negative, but due to the spectral folding the beat note will appear at positive frequency. Consequently, the related RF beat notes spaced by $\Delta f_\mathit{rep}$) will also be folded. An alternative view is that the lowest-frequency beat note undergoes a negative offset shift $ f_\mathit{off}$ (in counterphase) or that the zero-numbered beat note appears as second in the RF spectrum, assuming that the offset frequency is defined modulo the repetition rate~\cite{burghoff_computational_2016} and is constrained to be positive. 
	 	 
	 As a result, the mapping of the RF beat notes to the optical domain, which starts from the lowest frequency beat note, will cause a mirroring of the optical spectrum. This ambiguity and change of the mapping direction makes the assessment of the beat notes' sign of great importance.
	 
	 \begin{figure}[H]
	 	\centering
	 	\includegraphics[width=0.49\textwidth]{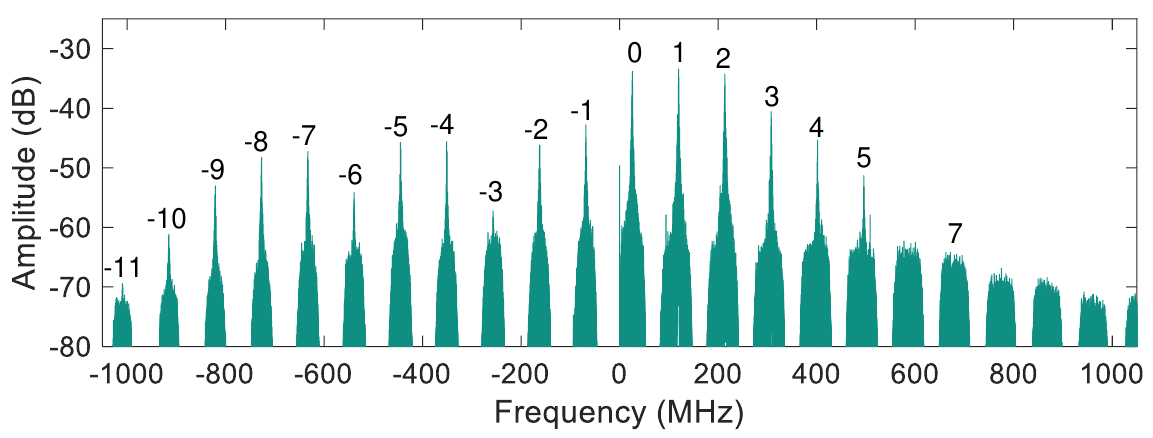}
	 	\caption{\textbf{Unfolded RF spectrum} as a result of bandpass filtering and phase inversion of the negative beat notes made possible after applying the Hilbert transform and combining the two parts. Note the harmonics of the repetition frequency at 94~MHz after the timing correction and the DC component peak. \label{fig:Fig8}}
	 \end{figure}   	 
	 
	 One way to distinguish the two cases of folding is to determine in which direction the lowest frequency beat note moves when the injection current of the sample-interacting laser is changed. This information can easily be obtained through a frequency discriminator feedback circuit, where the sign of the control signal will determine whether the lowest-frequency beat note is a result of optical beating lower or higher in frequency compared to the LO. Figure 5 shows a situation where the lowest frequency beat note is followed by a folded beat note, which is consistent with the established convention. 
	 
	 Having determined the beat note signs followed by selective multi-frequency bandpass filtering, we can extract time domain signals that contain only the positive, $y_p(t)$, or only the negative beat notes, $y_n(t)$, while the complete measured signal is a superposition of both complementary signals:
   	 \begin{equation}
   	 	 y(t) = y_p(t) + y_n(t) \;.
   	 \end{equation}
	 In order to enable negative frequency representation, the data need to be transformed into the complex domain. For this reason, the signals $y_p(t)$ and $y_n(t)$ are Hilbert-transformed into analytical form, which preserves only the positive frequencies of their Fourier domains. Finally, to unfold the spectrum a complex conjugate of the negative beat note analytic signal, $\tilde{y}_n^*(t)$ , is added to the positive,
   	 \begin{equation}
   	 \tilde{y}(t) = \tilde{y}_p(t) + \tilde{y}_n^*(t) \;.
   	 \end{equation}
	 Figure~\ref{fig:Fig8} shows the DC-centered complex frequency spectrum of $\tilde{y}_n^*(t)$, where the gaps between the subsequent beat notes are created via filtering and unfolding process. Notably, if the imaginary component of the signal is ignored, the real part should be identical to the non-processed $y(t)$. This conveniently allows for back-folding of the spectrum after the offset correction is performed, as described in the next subsection. 
	 
	 \subsection*{A.4 Frequency offset correction}
	 
	 The final step in the phase and timing correction for the multiheterodyne data is to correct for the offset frequency jitter that is common for all modes. After the timing correction, the beat note with the highest SNR and the narrowest line shape is used for the instantaneous phase fluctuation retrieval to suppress the noise contribution perturbing the estimate. 
	 
	 	 \begin{figure}[H]
	 	 	\centering
	 	 	\includegraphics[width=0.49\textwidth]{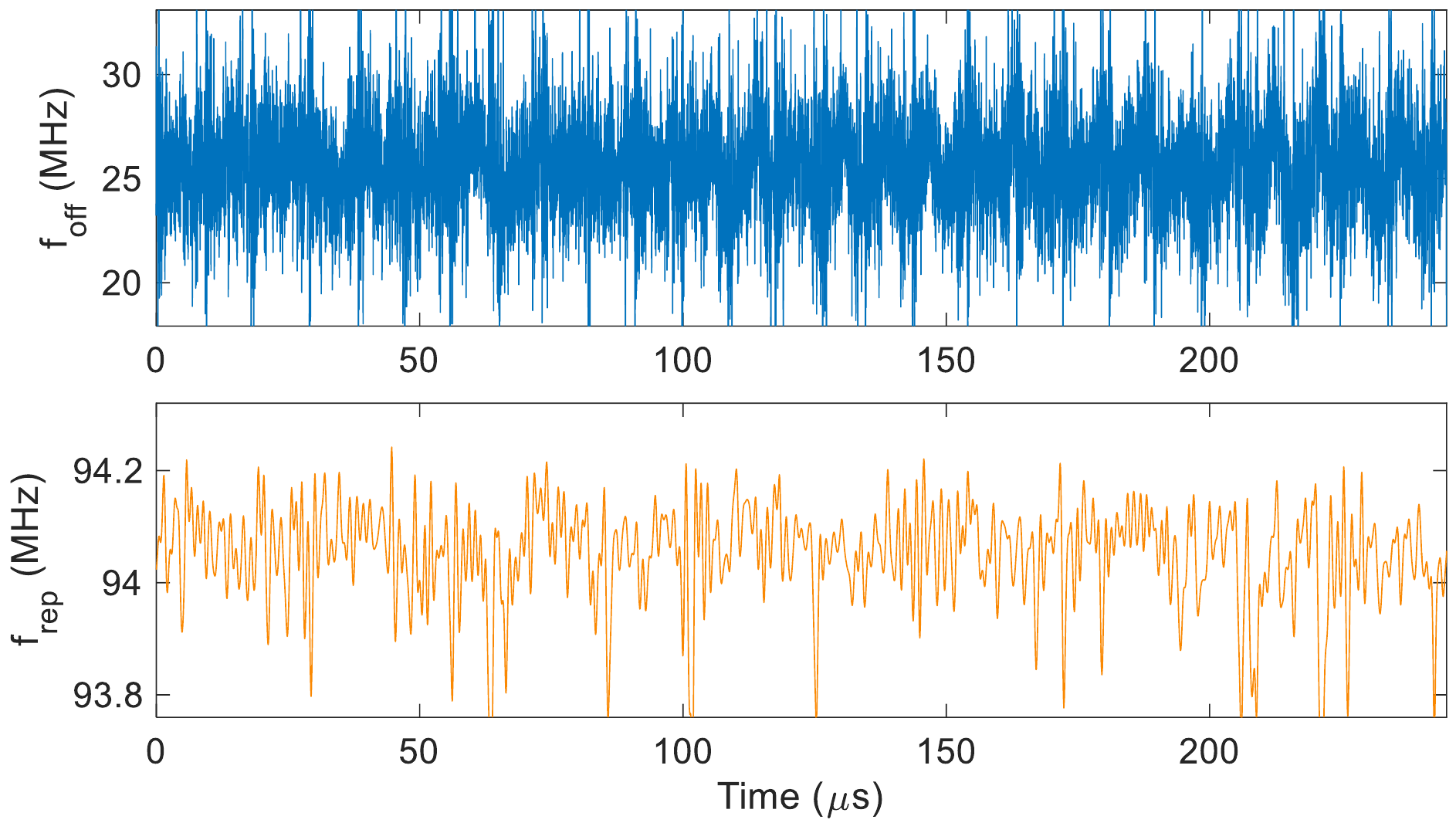}
	 	 	\caption{\textbf{Offset frequency $f_\mathit{off}$ and repetition rate frequency $\Delta f_\mathit{rep}$ fluctuations} with a 250~$\upmu$s acquisition time. The offset frequency fluctuations greatly exceed those of the repetition rate, causing mutually coherent multiheterodyne beat note phase noise.  \label{fig:Fig9}}
	 	 \end{figure} 	 
  	
	 The instantaneous phase, $\varphi_\mathit{off}$, is derived from the complex representation of the multiheterodyne signal after bandpass filtering with edge-effect suppression through data padding and phase linearization, as described before for the IIR filter. Most importantly, since this correction is performed directly on the instantaneous phase, the instantaneous frequency calculation that may introduce numerical artifacts caused by discrete numerical differentiation, is not needed. For the data in this work (including spectrally interleaved measurements), a bandpass filter bandwidth of 0.8$\Delta f_\mathit{rep}$/2 and a beat note SNR of 10-20 dB were sufficient for the $f_\mathit{off}$ retrieval. Figure~\ref{fig:Fig9} plots the retrieved instantaneous offset frequency as well as the instantaneous repetition rate. It is clear that in absolute terms the relative offset frequency, $f_\mathit{off}$, varies by approximately two orders of magnitude more than the repetition frequency difference, $\Delta f_\mathit{rep}$, which causes significant beat note broadening.

	 Finally, the relative offset frequency correction is performed by multiplying complex, spectrally-unfolded multiheterodyne signals by a linearly varying phase term corresponding to a constant offset frequency ($\varphi_\mathit{off}$~=~$f_\mathit{off}t$ when $f_\mathit{off}$ = const.). The constant offset frequency can be assumed based on either the mean value determined as the center position of the beat note used for the correction, the expected value of $f_\mathit{off}$ assumed a priori, or determined as a slope of a global linear fit to the instantaneous phase. 
	 The former linear fit method typically outperforms the other two methods, as it does not make any assumption regarding the frequency of the beat note, which is a source of potential errors. Once the offset frequency fluctuations are extracted from one beat note, it is applied to correct the phase of all multiheterodyne beat notes within the complex multiheterodyne signal. This eliminates the fluctuations of $f_\mathit{off}$ and ensures constant beat note frequencies throughout the acquisition. In other words, the retrieved instantaneous phase, $\varphi_\mathit{off}$, of the timing-corrected beat note centered at $\langle f_\mathit{off} \rangle$ is inserted into the equation 
	 \begin{equation}
	 \begin{split}
	 \tilde{y}_\mathit{corr}(t) &= exp\left(-i2\pi \int_0^t f_\mathit{off}(\tau)-\langle f_\mathit{off} \rangle d\tau \right) \cdot \tilde{y}(t) \\
	 &=exp\left( -i\varphi_\mathit{off}(t)+i2\pi \langle f_\mathit{off} \rangle t  \right) \cdot \tilde{y}(t) \;,
	 \end{split}
	 \end{equation}
	 where the first exponential term represents the phase-correction process.

\begin{figure}[H]
	           	\centering
	           	\includegraphics[width=0.49\textwidth]{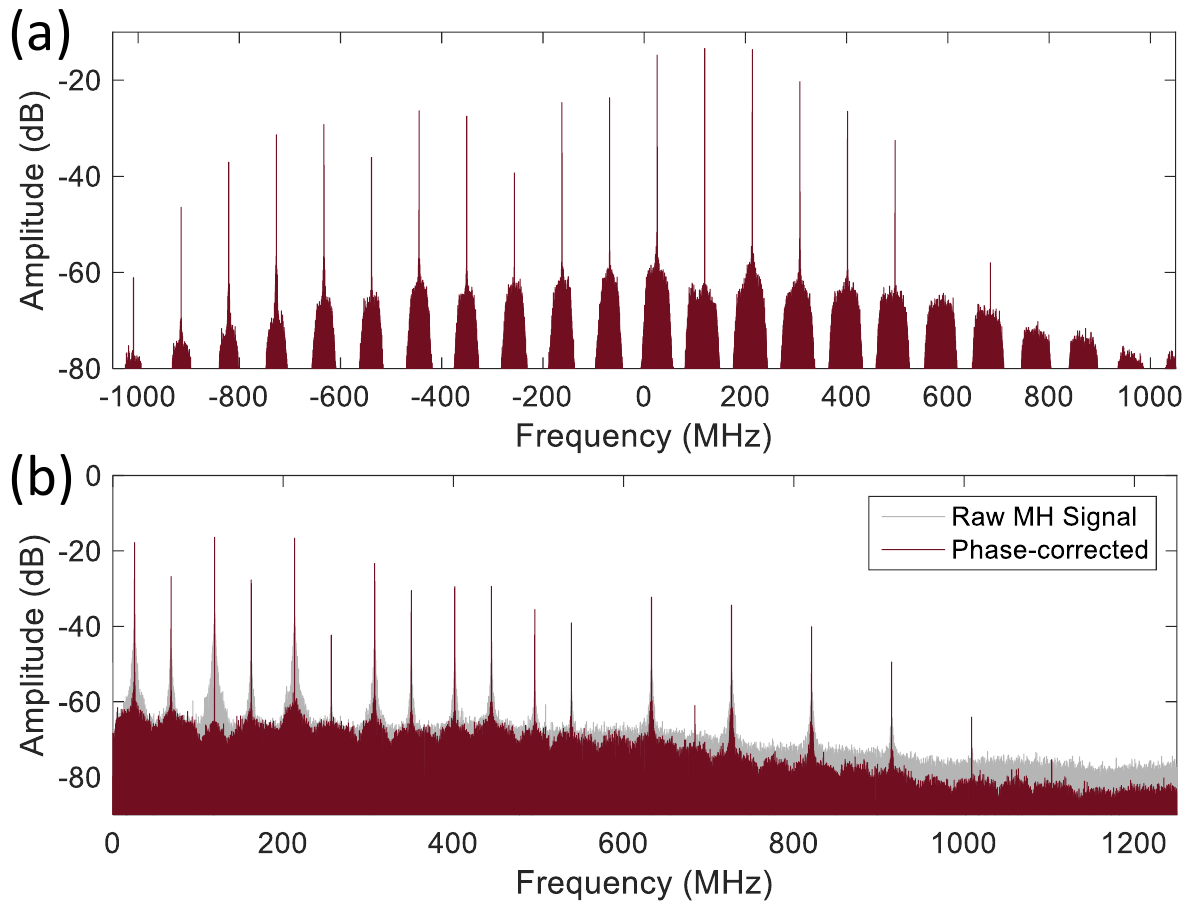}
	           	\caption{\textbf{RF spectrum after the computational timing and offset correction. (a)} Unfolded and corrected version of the original RF spectrum. \textbf{(b)} The gray trace shows the uncorrected original spectrum, whereas the back-folded and corrected spectrum is shown in red. Not only are the beat notes narrower with improved SNR, but also the noise floor is slightly lowered, which reveals weak beat notes close to the upper limit of the photodetector bandwidth.\label{fig:Fig10}}
	           \end{figure}

	 As shown in Fig.~\ref{fig:Fig10} this procedure leads to a significant linewidth reduction of the multiheterodyne beat notes, in this case fundamentally limited by the total acquisition time (2~kHz corresponding to 500~$\upmu s$ of acquisition time). This is a strong indication that the multiheterodyne beat notes are in fact coherent, albeit with considerable offset frequency jitter. If the opposite were true, a comb-based timing and phase correction procedure such as outlined here would only result in a broadening of the beat notes. 
	 
     \subsection*{Acknowledgments}
     We thank Dr. Adam Piotrowski of Vigo Systems for providing the optimized ultra-fast MCT detectors used in these studies. The work at Princeton was supported by the DARPA SCOUT program (W31P4Q161001) and the work at NRL was supported by ONR.
     
     
     \bibliographystyle{naturemag_noURL}
     \bibliography{ICL_MHS_references}
     
   \end{multicols}

\end{document}